\documentclass[aps,pra,reprint,nofootinbib,superscriptaddress,twocolumn,showpacs,showkeys,longbibliography,amsmath,amssymb]{revtex4-2}
\usepackage{graphicx}% Include figure files
\usepackage{dcolumn}% Align table columns on decimal point
\usepackage{bm}% bold math
\usepackage{braket}
\usepackage{subfigure}
\usepackage[colorlinks,bookmarks=false,citecolor=blue,linkcolor=red,urlcolor=blue]{hyperref}
\usepackage[english]{babel}
\usepackage{changes}

\begin{document}
\preprint{APS/123-QED}

\title{Nonuniversal Equation of State of a Quasi-2D Bose Gas in Dimensional Crossover}
\author{Xiaoran Ye}
\affiliation{Department of Physics, Zhejiang Normal University, Jinhua 321004, People's Republic of China}
\author{Tao Yu}
\affiliation{Department of Physics, Zhejiang Normal University, Jinhua 321004, People's Republic of China}
\author{Zhaoxin Liang}\email[Corresponding author:~] {zhxliang@zjnu.edu.cn}
\affiliation{Department of Physics, Zhejiang Normal University, Jinhua 321004, People's Republic of China}
\date{\today}% It is always \today, today, 	

\begin{abstract}
Equation of state (EOS) for a pure two-dimensional (2D) Bose gas exhibits a logarithmic dependence on the s-wave scattering length [\href{https://doi.org/10.1103/PhysRevLett.118.130402}{L. Salasnich, Phys. Rev. Lett. 118, 130402 (2017)}]. The pronounced disparity between the EOS of a 2D Bose gas and its 3D counterpart underscores the significance of exploring the dimensional crossover between these two distinct dimensions. In this work, we are motivated to deduce nonuniversal corrections to EOS for an optically trapped Bose gas along the dimensional crossover from 3D to 2D, incorporating the finite-range effects of the interatomic potential. Employing the framework of effective field theory, we derive the analytical expressions for both the ground state energy and quantum depletion. The introduction of the lattice induces a transition from a 3D to a quasi-2D regime. In particular, we systematically analyze the asymptotic behaviors of both the 2D and 3D aspects of the model system, with a specific focus on the nonuniversal effects on the EOS arising from finite-range interactions. The nonuniversal effects proposed in this study along the dimensional crossover represent a significant stride toward unraveling the intricate interplay between dimensionality and quantum fluctuations.
\end{abstract}
\maketitle
\section{Introduction}

Equation of state (EOS), representing a functional relationship among system variables, stands as a fundamental concept in elucidating quantum many-body systems. The universality of an EOS holds particular significance, enabling the description of diverse physical systems through a shared EOS. A paradigmatic illustration is the universal Lee-Huang-Yang (LHY) correction applied to EOS of weakly-interacting bosonic systems~\cite{leehy1957,leey1957}. In more details, a single parameter, the s-wave scattering length $a_{\text s}$, adeptly characterizes both the two-body problem and, consequently, the many-body physics.

In contrast, nonuniversal effects~\cite{andersen2004} in EOS are referred to the physical quantities that depend on parameters other than $a_{\text s}$. Recently, there has been a significant surge of interest in investigating these nonuniversal effects within the context of ultracold atomic gases. This heightened interest is motivated by the tuning capabilities of $a_{\text s}$, achieved through the use of magnetic~\cite{Chin2010} and optical~\cite{WuPRL2012} Feshbach resonances. Consequently, the finite-range parameter $r_{\text e}$, representing the next-to-leading order term in the interaction potential, cannot be disregarded. The nonuniversal effects induced by this parameter naturally arise.

Along this research line, the exploration of nonuniversal effects stemming from the finite range of the interatomic potential~\cite{salasnich2016,Mistakidis2023,Lorenzi2023} in ultracold atomic gases has garnered extensive attention. At the mean-field level, the nonuniversal corrections yield a modified Gross-Pitaevskii equation~\cite{Fu2003,Collin2007,QI2013,Veksler2014} governing the behavior of the nonuniform condensate. Extending beyond mean-field considerations, the thermodynamics induced by the finite-range interaction is derived up to the Gaussian level for both a pure 2D~\cite{salasnich2017} and 3D~\cite{cappellaro2017,tononi2018} uniform Bose gas. Notably, the non-trivial case of a 3D Fermi gas is examined in Refs.~\cite{Hu2020,Tajima2022,Sakakibara2023}. The existing contrast in nonuniversal effects between pure 2D~\cite{salasnich2017} and 3D~\cite{salasnich2016,tononi2018} scenarios adds significant depth to the study of systems existing between these distinct dimensions—a dimensional crossover, holding paramount fundamental interest. This work specifically aims to delve into the nonuniversal effects along the dimensional crossover from 3D to 2D in an optically-trapped Bose gas.

The dimensional crossover serves as a conduit for exploring diverse behaviors among systems of distinct dimensions~\cite{dyke2011,peppler2018,bloch2008,orso2006,lammers2016,orso2005,Li2022,Guo2024,Yao2023}. Existing facilities now allow for the tight confinement of trapped bosons in one direction, creating quasi-2D Bose gases. These gases exhibit kinematic 2D behavior, frozen in the confined direction~\cite{petrov2000,petrov2001,bisset2009,Hu2019,Yin2020,Jalm2019}, introducing a new length scale $a_{\text{2D}}$, which competes with the 3D scattering length $a_{\text{3D}}$~\cite{petrov2000,gorlitz2001}. Furthermore, optically trapped Bose gases provide enhanced experimental control, offering tunable interatomic interactions, adjustable tunneling amplitudes between adjacent sites, atom filling fractions, and lattice dimensionality~\cite{yukalov2009,zhou2010,burger2002,fort2003,rychtarik2004,Guo2023}. Specifically, under a 1D optical lattice, Bose gases undergo a dimensional crossover from 3D to quasi-2D, a phenomenon demonstrated both experimentally and theoretically~\cite{burger2002,hu2011,hu2013}. Given the capability to manipulate a 1D optical lattice and implement effective finite-range interactions, an intriguing avenue of inquiry involves examining how finite-range interaction influences a Bose gas confined within a 1D optical lattice.

In this work, employing effective field theory within the one-loop approximation, we obtain analytical expressions for the ground state energy and quantum depletion of a 1D-optically-trapped Bose gas with finite-range effective interaction at zero temperature. Exploiting the 3D to quasi-2D crossover induced by the introduction of the optical lattice, we scrutinize the impact of nonuniversal effects due to finite-range effective interaction in the asymptotic 2D regime. Our findings in the quasi-2D regime, incorporating finite-range interactions, exhibit a resemblance to the results observed in a homogeneous 2D Bose gas in Refs.~\cite{salasnich2017,tononi2018}. When the finite-range interaction diminishes, our results are consistent with those presented in Refs.~\cite{schick1971,andersen2002,zhou2010}. Thus, our results of EOS provide a substantial bridge for understanding the pronounced disparity between the logarithmic dependence of EOS on finite-range parameters in the 2D case and its 3D counterpart. We also remark that the strategy of realizing dimensional crossover from full 3D to quasi-low-D can go beyond the scope of the current work and be extended into other physical systems in the context of the ultracold quantum gas, e.g. the dipolar BEC system~\cite{Boudjem2013,Mishra2020}, Bose-Bose mixture~\cite{Li2023} or the quantum droplet~\cite{Petrov2015,Petrov2016}.

The paper is structured as follows. In Sec.~\ref{MTTBG}, we present the Hamiltonian of the model system and outline the basic framework of the path integral formalism. In Sec.~\ref{GSEAQD}, the analytical expressions of the ground state energy and quantum depletion are calculated using the effective field theory within the one-loop approximation. Sec.~\ref{DCB} offers a comprehensive examination of the influence of dimensional crossover on ground-state properties induced by a 1D optical lattice. The nonuniversal effects in the dimensional crossover regimes are discussed. Sec.~\ref{CCS} summarizes our work, and we also consider the conditions for the potential experimental realization of our scenario.

\section{Model system and Hamiltonian}\label{MTTBG}
In this work, we consider a 3D Bose-Einstein Condensation (BEC), accounting for the finite-range effects of the interatomic potential in the following geometry: in the $x$ direction, the BEC is trapped in an optical lattice, while in the $y$ and $z$ directions, the atoms are free. Such a physical system can be well described within the framework of the path integral formalism. In more details, our starting point is the grand-canonical partition function of a 3D interacting dilute Bose gas in the presence of a 1D optical lattice~\cite{atland2010}
\begin{eqnarray}
	\mathrm{Z} & = & \int D\left[\psi^{*},\psi\right]\exp\left\{-\frac{S\left[\psi^{*},\psi\right]}{\hbar}\right\}.\label{Pf}
\end{eqnarray}
with the action functional $S\left[\psi^{*},\psi\right]$ in Eq.~(\ref{Pf}) reading~\cite{cappellaro2017,salasnich2017,tononi2018}
\begin{widetext}
\begin{eqnarray}
S=  \int_{0}^{\hbar\beta}d\tau\int d^3\mathbf{r} \psi^{*}\left(\mathbf{r},\tau\right)\left[\hbar\frac{\partial}{\partial\tau}-\frac{\hbar^{2}\nabla^{2}}{2m}-\mu+V_{\text{opt}}\left(\mathbf{r}\right)\right]\psi\left(\mathbf{r},\tau\right) +\frac{g_{0}}{2}\left|\psi\left(\mathbf{r},\tau\right)\right|^{4}-\frac{g_{2}}{2}\left|\psi\left(\mathbf{r},\tau\right)\right|^{2}\nabla^{2}\left|\psi\left(\mathbf{r},\tau\right)\right|^{2}.
	\label{Lg}
\end{eqnarray}
\end{widetext}
In Eq.~(\ref{Lg}), $\psi\left(\mathbf{r},\tau\right)$, describing the atomic bosons, represents the complex field in both space ${\bf r}$ and imaginary time $\tau$. Here, $\beta\equiv1/(k_{\text B}T)$, where $k_{\text B}$ is the Boltzmann constant and $T$ is the temperature. Additionally, $\mu$ denotes the chemical potential. The parameters $g_0$ and $g_2$ correspond to the effective two-body coupling constants in the presence of a 1D optical lattice. These constants are associated with the two-body s-wave scattering length~\cite{dalfovo1999,haugset1998,roth2001} and the s-wave effective range~\cite{braaten2001,cappellaro2017}, respectively.

The $V_{\text{opt}}\left(\mathbf{r}\right)$ represents the 1D optical lattice, reading~\cite{bloch2008}
\begin{eqnarray}
	V_{\text{opt}}\left(\mathbf{r}\right) & = & sE_{\text R}\sin^{2}\left(q_{\text B}x\right),\label{Lp}
\end{eqnarray}
In Eq.~(\ref{Lp}), the $s$ denote the intensity of a laser beam, $E_{\text R}=\hbar^{2}q_{\text B}^{2}/2m$ is the recoil energy, with $\hbar q_{\text B}$ being the Bragg mometum and $m$ the atomic mass. The lattice period is fixed by $q_{\text B}=\pi/d$ with $d$ being the lattice spacing. Atoms are free in the $y-z$ plane. 

Before delving into the exploration of nonuniversal effects along the dimensional crossover from 3D to quasi-2D based on Eq.~(\ref{Pf}), we initially provide a brief overview of key features of a BEC with finite-range effective interaction in uniform space, corresponding to Eq.~(\ref{Lg}) with $V_{\text{opt}}=0$. It's worth noting that the nonuniversal equation of state for the uniform Bose gas with finite-range effects has been previously derived in both pure uniform 3D and 2D cases using effective field theory~\cite{cappellaro2017,salasnich2017}.

Introducing an additional optical lattice ($V_{\text{opt}}\neq 0$) to the aforementioned BEC in uniform space introduces hierarchical access to new energy and length scales, consequently inducing a dimensional crossover from 3D to low-D~\cite{zhou2010}. In more details, by controlling the depths of the optical lattice $V_{\text{opt}}\left(\mathbf{r}\right)$, dimensional crossovers to lower dimensions are anticipated to occur in the following manner: a 3D Bose gas transitions to quasi-2D when the energetic constraint to freeze $x$-direction excitations is reached.

It's important to note that the tight confinement in the direction of the optical lattice significantly influences the value of the effective coupling constant in Eq.~(\ref{Lg}). Specifically, in the presence of the optical lattice, both the s-wave coupling constant $g_0$ and the finite-range coupling constant $g_2$ generally exhibit dependence on density and lattice parameters. This stands in marked contrast to a free 3D Bose gas, where $g_0^{\text{3D}} = 4\pi\hbar^2a_{\text{3D}}/m$, and $g_2^{\text{3D}} = 2\pi\hbar^2a_{\text{3D}}r_{\text e}/m$ , with $a_{\text{3D}}$ and $r_{\text e}$ representing the 3D scattering length and the finite range constant, respectively. For the sake of formulation clarity, however, we will use $g_0$ and $g_2$ for notational convenience while temporarily setting aside their specific expressions to derive general formulations for the ground-state energy and quantum depletion. Finally, we remark that the effects of confinement-induced resonance (CIR)~\cite{Peng2010,Zhang2011} on the coupling constant are not considered here. The key physics of CIR can be captured through the language of Feshbach resonance~\cite{Bergeman2003}, where the scattering open channel and closed channels are represented by the ground-state transverse mode and other transverse modes along the tight-confinement dimensions, respectively. Under the tight-binding approximation assumed in this work, ultracold atoms are frozen in the states of the lowest Bloch band and cannot be excited into the other transverse modes. Therefore, the effect of CIR on the coupling constant can be safely ignored as the closed channels are absent.

\section{Nonuniversal equation of state of model system}\label{GSEAQD}

\begin{figure}[t]
\begin{centering}
\includegraphics[scale=0.7]{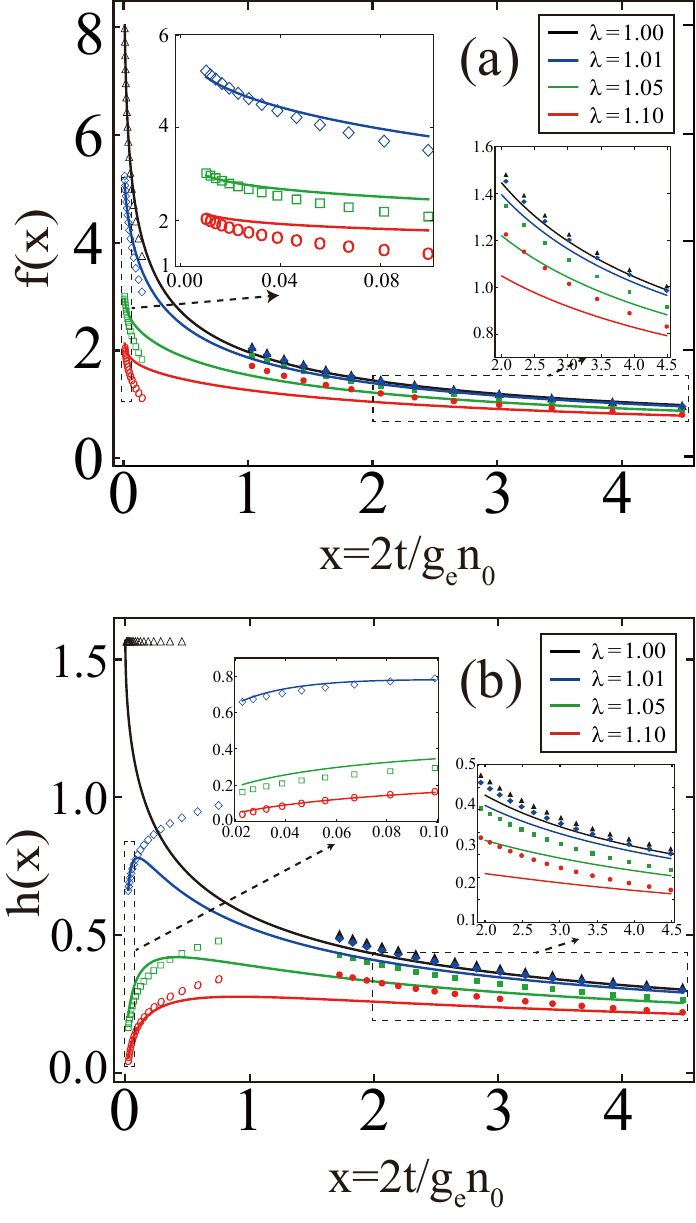}
\par\end{centering}
\caption{(a) Scaling function $f(x)$ in Eq.~(\ref{Fx}) for different values of $\lambda=1+\frac{4m\mu}{\hbar^{2}}\frac{g_{2}}{g_{0}}$. Here, the solid- and empty-point curves denote the 3D and quasi-2D asymptotic behaviors of $f(x)$ respectively. (b) Scaling function $h(x)$ in Eq.~(\ref{hx}) for different values of $\lambda$. The solid- and empty-point curves denote the 3D and quasi-2D asymptotic behaviors of $h(x)$ respectively.\label{1}}
\end{figure}

In what follows, we focus on the situation where the optical lattice is strong enough to create many separated wells, giving rise to an array of condensates, while maintaining full coherence through quantum tunneling. With this assumption, one can refer to $n_0$ as the condensate density and safely neglect the Mott insulator phase transition. Under these conditions~\cite{zhou2010}, it becomes possible to investigate the nonuniversal EOS of the model system using effective field theory.

In this work, we restrict ourselves to the case where the laser intensity $s$ is relatively large that the chemical potential $\mu$ is small compared to the interwell barriers, thus we only consider the lowest Bloch band~\cite{orso2006}. In the tight-binding approximation, the condensate of the lowest Bloch band can be written in terms of Wannier function as $\phi_{k_{x}}(x)=\sum_{l}e^{ildk_{x}}w\left(x-ld\right)$, where $w\left(x\right)=d^{1/2}\exp\left[-x^{2}/2\sigma^{2}\right]/\pi^{1/4}\sigma^{1/2}$ with $d/\sigma\simeq\pi s^{1/4}\exp\left(-1/4\sqrt{s}\right)$. 

We proceed to expand the bosonic complex field of the action in Eq.~(\ref{Lg})  as follows 
\begin{eqnarray}
\psi\left(\mathbf{r},\tau\right)  =  \sum_{\mathbf{k},n}{\psi}_{\mathbf{k},n}\phi_{k_{x}}(x)e^{-i(k_{y}y+k_{z}z)}e^{i\omega_{n}\tau},\label{CPF}
\end{eqnarray}
with $\omega_{n}=2\pi n/\hbar\beta$ being the bosonic Matsubara frequencies and $n$ being integers. By plugging Eq.~(\ref{CPF}) into  Eq.~(\ref{Lg}),  we can rewrite the action of model system with the following form (see Appendix~\ref{AppendixA} for the detailed derivation)
\begin{eqnarray}
\!\!&&\frac{S[\psi_{\mathbf{k},n}^{*},\psi_{\mathbf{k},n}]}{\hbar\beta V}=\sum_{\mathbf{k},n}\psi_{\mathbf{k},n}^{*}(-i\hbar\omega_{n}+\varepsilon^0_{\mathbf{k}}-\mu)\psi_{\mathbf{k},n}\nonumber\\
 \!\!\!\!&&+\!\!\sum_{\substack{\mathbf{k,k^{\prime},q}\\n,n^{\prime},m}}\!\!(\frac{g_e}{2}\!\!+\!\!\frac{g_{2}\tilde{q}^2}{2}\!)\psi_{\mathbf{k+q},n+m}^{*}\psi_{\mathbf{k^{\prime}-q},n^{\prime}-m}^{*}\psi_{\mathbf{k^{\prime}},n^{\prime}}\psi_{\mathbf{k},n},\label{EFunctional}
\end{eqnarray}
where $V$ is the volume of the system and $\varepsilon^0_{\mathbf{k}}=2t(1-\cos k_{x}d)+\hbar^2k_\perp^2/2m$ represents the energy dispersion of the noninteracting model with $t=-(1/d)\int^{d}_{0}{dx}w(x)\left(-\hbar^{2}\partial_{x}^{2}/2m+V_{\text{opt}}\right)w(x-d)$ being the tunneling rate along the $x$-direction between neighboring wells.

The functional~(\ref{EFunctional}) can be utilized to calculate the nonuniversal EOS of an optically-trapped Bose gas with finite-range interaction. Comparing this functional to the one characterizing a Bose gas without optical confinement~\cite{cappellaro2017,salasnich2017}, two significant differences arise with the introduction of an optical lattice: First, the kinetic energy term along the $x$-direction (denoted as $\varepsilon^0_{\mathbf{k}}$ in functional~(\ref{EFunctional})) no longer assumes the classical quadratic form present in the radial direction. Instead, it exhibits a periodic band structure. In the limit of $4t\gg \mu$, the system maintains an anisotropic 3D behavior with $\varepsilon^0_{\mathbf{k}}=\hbar^2k_x^2/2m^*+\hbar^2(k_y^2+k_z^2)/2m$ and $m^*=\hbar^2/2td^2$ being the effective mass associated with the band. Conversely, for $4t\ll \mu$, the system undergoes a dimensional crossover to a 2D regime when the energetic restriction to freeze axial excitations is reached with $\varepsilon^0_{\mathbf{k}}\simeq \hbar^2k_\perp^2/2m$. Second, the s-wave interaction coupling constant $g_0$ is renormalized to $g_{\text e}=g_{0}d/\sqrt{2\pi}\sigma$ due to the presence of the optical lattice. Remarkably, regularization of the s-wave finite-range interaction coupling constant, influenced by restricted kinematics, introduces a new dimensional crossover from 3D to 2D in the context of interaction energy. In more details, the s-wave finite-range interaction coupling constant in the last term of the functional~(\ref{EFunctional}) can be rewritten as 
\begin{eqnarray}
	\frac{g_2\tilde{q}^2}{2}=\frac{2g_2m}{\hbar^2}\times \left[\frac {\hbar^2}{2m}t_{1}(1-\cos k_xd)+\frac {\hbar^2k_\perp^2}{2m(\sqrt{8\pi}\sigma/d)}\right],\label{EFS}
\end{eqnarray}
with $ \hbar^2t_{1}/2m=(1/d)\int ^{d}_{0}dxw^{2}(x)[\hbar^2\partial_{x}^{2}/2m]w^{2}(x-d)$. 
Consequently, from accessing one extreme of $\hbar^2t_{1}/2m/\mu$ to the other in Eq.~(\ref{EFS}), two distinctive regimes related to the s-wave finite range interaction coupling constant are further identified and another dimensional crossover emerges.

By applying the Gaussian (one-loop) approximation to the action~(\ref{EFunctional}) by writing $\psi_{\mathbf{k},n}=\psi_{0,n}+\eta\left(\mathbf{k},i\omega_{n}\right)$ and proceeding in the standard fashion~\cite{cappellaro2017,salasnich2017}, one can obtain the Gaussian contribution of quantum fluctuation to the action~(\ref{EFunctional}) as follows
\begin{eqnarray}
S_{\text{eg}}=  \frac{1}{2}\sum_{Q}(\tilde{\eta}^{*}\left(Q\right),\tilde{\eta}\left(-Q\right))\mathbf{M}\left(Q\right)\left(\begin{array}{c}
\tilde{\eta}\left(Q\right)\\
\tilde{\eta}^{*}\left(-Q\right)
\end{array}\right),
\end{eqnarray}
with $Q=\left(\mathbf{k},i\omega_{n}\right)$ is the $3+1$ vector denoting the momenta $\mathbf{k}$ and bosonic Matsubara frequencies $\omega_{n}$, and the inverse fluctuation propagator $\mathbf{M}\left(Q\right)$ reads
\begin{widetext}
\begin{eqnarray}
\!\!\!\mathbf{M}=\beta\left(\begin{array}{cc}
\!\!-i\hbar\omega_{n}\!+\!\varepsilon^0_{\mathbf{k}}\!-\!\mu\!+\!\left\{2g_{\text e}\!+\!\frac{2g_2m}{\hbar^2} \left[\frac {\hbar^2t_{1}(1\!-\!\cos k_xd)}{m}\!+\!\!\frac {\hbar^2k_\perp^2}{2m(\sqrt{2\pi}\sigma/d)}\right]\right\}n_0,\!\!\!\!\!\!\!\!\!\!\!\! & \!\!\!\!\!\!\!\left\{g_{\text e}+\frac{2g_2m}{\hbar^2} \left[\frac {\hbar^2t_{1}(1-\cos k_xd)}{m}+\frac {\hbar^2k_\perp^2}{2m(\sqrt{2\pi}\sigma/d)}\right]\right\}n_0\\
\!\!\!\!\!\!\!\!\!\!\!\!\!\!\!\!\!\!\!\!\!\!\!\!\left\{g_{\text e}+\frac{2g_2m}{\hbar^2} \left[\frac {\hbar^2(1-\cos k_xd)}{m}t_{1}+\frac {\hbar^2k_\perp^2}{2m(\sqrt{2\pi}\sigma/d)}\right]\right\}n_0, \!\!\!\!\!\!\!\!\!\!\!\!& \!\!\!\!\!\!\!\!\!\!\!\!i\hbar\omega_{n}\!+\!\varepsilon^0_{\mathbf{k}}\!-\!\mu\!+\!\left\{2g_{\text e}\!+\!\frac{2g_2m}{\hbar^2} \left[\frac {\hbar^2t_{1}(1\!-\!\cos k_xd)}{m}\!+\!\frac {\hbar^2k_\perp^2}{2m(\sqrt{2\pi}\sigma/d)}\right]\right\}n_0
\end{array}\right).
\label{Propagator}
\end{eqnarray}
\end{widetext}

Before proceeding with further calculations, we double-check whether Eq.~(\ref{Propagator}) can be simplified into the existing previous results when either the optical lattice, $V_{\text{opt}}$, vanishes, or the finite-range interaction, $g_2$, vanishes, or both. In more details, in the limit of $g_2=0$ and $V_{\text{opt}}=0$, Eq.~(\ref{Propagator}) should recover the corresponding one in Ref.~\cite{cappellaro2017} (refer to Eq.~(9) in Ref.~\cite{cappellaro2017}). Next, in the limit of $g_2\neq 0$ and $V_{\text{opt}}=0$, our result in Eq.~(\ref{Propagator}) should align with the corresponding one in Ref.~\cite{cappellaro2020} (see Eq.~(15) in Ref.~\cite{cappellaro2020}). Then, in the limit of $g_2=0$ and $V_{\text{opt}}\neq 0$, our result for Eq.~(\ref{Propagator}) should be consistent with the corresponding one in Ref.~\cite{zhou2010}.

We proceed to integrate over the bosonic fields of action~(\ref{EFunctional}) and obtain the Gaussian grand potential 
\begin{eqnarray}
\Omega_{\text{g}}=\frac{1}{2\beta}\sum_{Q}\ln{\det}[M(Q)]=\sum_{\bf k}\left(\frac{E_{\bf k}}{2}+\frac{1}{\beta}\ln\left(1-e^{-\beta E_{\bf k}}\right)\right),\label{GGP}
\end{eqnarray}
with $E_{\bf k}$ being the excitation energy of an optically-trapped Bose gas, reading 
\begin{eqnarray}
E_{\bf k}& = &\sqrt{\varepsilon^0_{\mathbf{k}}\left[\frac{\hbar^{2}k_{\perp}^{2}}{2m/(1+\frac{4m\mu}{\hbar^2}\frac{g_2}{g_0})}+\frac{\hbar^2(1-\cos k_xd)}{\hbar^2/[2\left(t+2g_2n_0t_1\right)]}+2\mu\right]},\nonumber \\
\end{eqnarray}
with the chemical potential being $\mu=g_en_0$. 

In this work, our focus lies on the nonuniversal EOS of the model system at zero temperature. 
To determine this, the ground state energy of the model system, which can be calculated from the zero-temperature grand potential $\Omega^{(0)}$ using the thermodynamic formula $E_g = \Omega^{(0)} + V\mu n_0$, is as follows: (detailed calculation can be found in Appendix~\ref{AppendixB})

\begin{widetext}
\begin{eqnarray}
\frac{E_{\text g}}{V} =\frac{1}{2}g_{\text e}n_{0}^{2}&+&\frac{1}{2V}\sum_{\mathbf{k}\ne0}\Big\{E_{\mathbf{k}}-\sqrt{\Big[\frac{\hbar^2k_\perp^2}{2m}+2t(1-\cos k_{x}d)\Big]\Big[\frac{\hbar^{2}k_{\perp}^{2}}{2m/(1+\frac{4m\mu}{\hbar^2}\frac{g_2}{g_0})}+\frac{\hbar^2(1-\cos k_xd)}{\hbar^2/[2\left(t+2g_2n_0t_1\right)]}\Big]}\nonumber\\
&-&\frac{g_{e}n_{0}\sqrt{\frac{\hbar^{2}k_{\perp}^{2}}{2m}+2t(1-\cos k_{x}d)}}{\sqrt{\frac{\hbar^{2}k_{\perp}^{2}}{2m/(1+\frac{4m\mu}{\hbar^{2}}\frac{g_{2}}{g_{0}})}+\frac{\hbar^{2}(1-\cos k_{x}d)}{\hbar^{2}/\left[2\left(t+2g_{2}n_{0}t_{1}\right)\right]}}}+\frac{(g_{e}n_{0})^2\sqrt{\frac{\hbar^{2}k_{\perp}^{2}}{2m}+2t(1-\cos k_{x}d)}}{2\left(\frac{\hbar^{2}k_{\perp}^{2}}{2m/(1+\frac{4m\mu}{\hbar^{2}}\frac{g_{2}}{g_{0}})}+\frac{\hbar^{2}(1-\cos k_{x}d)}{\hbar^{2}/[2\left(t+2g_{2}n_{0}t_{1}\right)]}\right)^{3/2}}\Big\},
\label{GSE}
\end{eqnarray}
\end{widetext}

In Eq.~(\ref{GSE}), the first term on the right side of the equality sign represents the mean-field contribution, while all the subsequent terms correspond to the corrections beyond the mean-field due to quantum fluctuations. It is noteworthy that the last two terms in Eq.~(\ref{GSE}) are introduced to circumvent ultraviolet divergence by employing an appropriate renormalization of the coupling constant~\cite{zhou2010,braaten1997}. In the continuum limit, we can systematically replace the summation in Eq.~(\ref{GSE}) with an integral and derive the analytical expression for the ground state energy of the model system as follows
\begin{eqnarray}
\frac{E_{\text g}}{V} & = & \frac{1}{2}g_{\text e}n_{0}^{2}+\frac{m\left(g_{\text e}n_{0}\right)^{2}}{\left(2\pi\right)^{2}\hbar^{2}d}f(x),\label{GSEC}
\end{eqnarray}
where the scaling function $f(x)$ in terms of the variable $x=2t/g_{\text e}n_{0}$ is defined as
\begin{eqnarray}
f(x) & = & \frac{1}{2}\int_{-\pi}^{\pi}dk_{x}^\prime \int_{0}^{\infty}dk\Big\{-\sqrt{\left(k+x\gamma\right)}\sqrt{k\lambda+\left(x+2t_{2}\right)\gamma}\nonumber \\
 &  & +\frac{\sqrt{\left(k+x\gamma\right)}}{2\left(k\lambda+\left(x+2t_{2}\right)\gamma\right)^{3/2}}-\frac{\sqrt{\left(k+x\gamma\right)}}{\sqrt{k\lambda+\left(x+2t_{2}\right)\gamma}}\nonumber \\
 &  & +\sqrt{\left(k+x\gamma\right)\left(k\lambda+\left(x+2t_{2}\right)\gamma+2\right)}\Big\},\label{Fx}
\end{eqnarray}
with $k_x^\prime=k_xd$ and $k$ being the dimensionless quasi-momentum and $\gamma=1-\cos k_x^\prime$ and $t_2=2g_2t_1/g_e$ and $\lambda=1+4mg_{2}n_0d/\sqrt{2\pi} \sigma \hbar^2$.
Eq.~(\ref{Fx}) can be numerically calculated, and the corresponding results are presented in Fig.~\ref{1}(a). Before delving into further analysis, we aim to verify the validity of Eq.~(\ref{GSEC}) by demonstrating that it can recover well-known results in the limiting cases. In the scenario of a vanishing optical lattice, i.e., $V_{{\text{opt}}}=0$, our model system simplifies into the pure 3D case. The ground-state energy of Eq.~(\ref{GSEC}) can then be reduced to
\begin{eqnarray}
	\frac{E_{\text g}}{V} & = & \frac{1}{2}g_{0}n_{0}^{2}+\frac{\left(ng_{0}\right)^{5/2}}{4\pi^{2}}\left(\frac{2m}{\hbar^{2}}\right)^{3/2}f_{0}(x),\label{G13D}
\end{eqnarray}
where the function of $f_{0}(x)$ reads
\begin{eqnarray}
	f_{0}(x) & = & \int_{0}^{\infty}dK_{0}K_{0}^{2}\Big(\sqrt{K_{0}^{2}\left(\lambda K_{0}^{2}+2\right)}-\sqrt{\lambda}K_{0}^{2}\nonumber \\
	&  & -\frac{1}{\sqrt{\lambda}}+\frac{1}{2\lambda^{3/2}K_{0}^{2}}\Big),
	\label{F0}
\end{eqnarray}
which can be solved analytically, yielding the result $f_{0}(x)=8\sqrt{2}/15\lambda^{2}$. Consequently, our result in Eq.~(\ref{G13D}) can precisely recover the relevant ones in Ref.~\cite{tononi2018,cappellaro2017}. Meanwhile, Eq.~(\ref{GSE}) is consistent with the one in Ref.~\cite{zhou2010} in the case of vanishing finite-range interaction, i.e., $g_2=0$.

We proceed to calculate the quantum depletion of the model system. The zero-temperature total particle number $N$ can be derived from the zero temperature grand potential $\Omega^{\left(0\right)}$ using the thermodynamic formula $N= -\partial\Omega^{\left(0\right)}/\partial\mu$, Consequently, the quantum depletion of the model system can be directly obtained as follows:
\begin{eqnarray}
\frac{N-N_{0}}{N} & = & \frac{mg_{\text e}}{2\pi^{2}\hbar^{2}d}h\left(x\right),\label{Qdeple}
\end{eqnarray}
where the function of $h\left(x\right)$ is defined as 
\begin{eqnarray}
h\left(x\right) & = & -\frac{1}{4}\int_{0}^{\infty}dk\int_{-\pi}^{\pi}d k_{x}^\prime\Big\{\frac{\sqrt{k+x\gamma}}{2\sqrt{k\lambda+(x+2t_{2})\gamma}}\nonumber \\
 &  & +\frac{\sqrt{k\lambda+(x+2t_{2})\gamma}}{2\sqrt{k+x\gamma}}-\frac{\sqrt{k+x\gamma}}{2(k\lambda+(x+2t_{2})\gamma)^{3/2}}\nonumber \\
 &  & -\frac{2+(\lambda+1)k+2x\gamma +2\gamma t_{2}}{2\sqrt{(k+x\gamma)(2+k\lambda+\gamma(x+2t_{2}))}}\nonumber \\
 &  & +\frac{1}{2\sqrt{k+x\gamma}\sqrt{k\lambda+(x+2t_{2})\gamma}}\Big\}.\label{hx}
\end{eqnarray}
which can be calculated numerically and the result is shown in Fig.~\ref{1}(b). 

Then, we routinely check the validity of Eq.~(\ref{Qdeple}) by using the limiting result of Eq.~(\ref{Qdeple}) in the case of vanishing lattice potential. By setting $V_{{\text{opt}}}=0$, the quantum depletion in Eq.~(\ref{Qdeple}) takes the following form:
\begin{eqnarray}
\frac{N-N_{0}}{N} & = & \frac{n^{1/2}}{(2\pi)^{2}}\left(\frac{2mg_{0}}{\hbar^{2}}\right)^{3/2}h_{0}\left(x\right),
\end{eqnarray}
with the function of $h_{0}(x)$ being
\begin{eqnarray}
h_{0}(x) & = & -\int_{0}^{\infty}K_{0}^{2}dK_{0}\Big[-\frac{K_{0}^{2}\lambda+K_{0}^{2}+2}{2\sqrt{K_{0}^{2}(K_{0}^{2}\lambda+2)}}\nonumber \\
 &  & +\frac{1}{2}\lambda^{-1/2}(\lambda+1)+\frac{2(\lambda-1)}{4K_{0}^{2}\lambda^{3/2}}\Big],
\end{eqnarray}
which can also be solved and the result is $h_{0}(x)=\sqrt{2}(\lambda-2)/3\lambda^{2}$,
which is exactly consistent with the corresponding result in Ref.~\cite{tononi2018}. 

Eqs.~(\ref{GSE}) and~(\ref{Qdeple}) stand as key results of this work, representing the nonuniversal EOS of an optically-trapped Bose gas with the finite-range effects of the interatomic potential. In the subsequent analysis, we intend to utilize Eqs.~(\ref{GSE}) and~(\ref{Qdeple}) to examine nonuniversal corrections to both the ground-state energy and quantum depletion of the model system along the dimensional crossover from 3D to quasi-2D.

\section{Nonuniversal Effects along dimensional crossover from 3D to quasi-2D}\label{DCB}

In the preceding Sec.~\ref{GSEAQD}, we have derived the analytical expressions for the EOS of an optically-trapped Bose gas with finite-range interaction. Building upon Eqs.~(\ref{GSE}) and~(\ref{Qdeple}), the purpose of Sec.~\ref{DCB} is to analyze the nonuniversal effects due to the finite-range interaction on the EOS along the dimensional crossover from 3D to quasi-2D. Dimensional crossovers are characterized by hierarchical access to new energy and length scales.  Based on Eq.~(\ref{EFunctional}), two kinds of dimensional crossover related to the kinetic and interaction energies respectively can be identified as shown in what follows. 

Firstly, the excitations of the model system can be frozen in the $x$-direction by the introduction of an optical lattice. In more details,  the $\varepsilon^0_{\mathbf{k}}=2t(1-\cos k_{x}d)+\hbar^2k_\perp^2/2m$ in Eq.~(\ref{EFunctional}) represents the the lattice-modified kinetic energy of the model system. Here, the $t$ is the tunneling rate along the $x$-direction between neighboring wells and is supposed to decay exponentially with the increasing the lattice depth. As a result, the concrete forms of the kinetic energy in Eq.~(\ref{EFunctional}) will change from the 3D form of $\varepsilon^0_{\mathbf{k}}=\hbar^2k^2/2m^*+\hbar^2k_\perp^2/2m$ to the 2D form of $\varepsilon^0_{\mathbf{k}}=\hbar^2k_\perp^2/2m$ as $t$ decreasing to be zero. 

Secondly, the interaction term in Eq.~(\ref{EFunctional}) exhibits 2D features of particle motion, and a dimensional crossover from 3D to 2D emerges in the behavior of the interaction energy when the energetic restriction to freeze axial excitations is reached.

(i) For $4t/\mu \gg 1$, the system exhibits an anisotropic 3D behavior, and the s-wave effective coupling constant takes form $g_{\text e}=\tilde{g}_{\text {3D}}= 4\pi\hbar^{2}\tilde{a}_{\text{3D}}/m$ with the lattice-renormalized $s$-wave scattering length $\tilde{a}_{\text{3D}}=a_{\text{3D}}d/(\sqrt{2\pi}\sigma)$. Furthermore,  the s-wave finite-range interaction coupling constant in the last term of the functional~(\ref{EFunctional}) takes the form of $\frac{g_2\tilde{q}^2}{2}=\frac{2g_2m}{\hbar^2}\times \left[\frac {\hbar^2k_x^2}{2m^*}+\frac {\hbar^2k_\perp^2}{2m(\sqrt{8\pi}\sigma/d)}\right]$. 

(ii) For $4t/\mu\ll 1$,  the two interacting bosons are in the ground state of an effective harmonic potential with a defined frequency $\omega_{0}=\hbar/m\sigma^{2}$ and harmonic oscillator length $\sigma$. The system undergoes a crossover to the quasi-2D regime, where the s-wave coupling constant is reduced to that in a tightly confined harmonic trap, given by $g_{\text{e}}=g_{\text{h}}d$~\cite{orso2006,petrov2000,zhou2010,petrov2001,hu2011} .
\begin{eqnarray}
g_{\text h} & = & \frac{2\sqrt{2\pi}\hbar^{2}}{m}\frac{1}{a_{\text{2D}}/a_{\text {3D}}+\left(1/2\pi\right)\ln\left[1/n_{\text{2D}}a_{\text{2D}}^{2}\right]},\label{Gh}
\end{eqnarray}
with the surface density $n_{\text{2D}}=n_{0}d$ and the effective 2D scattering length $a_{\text{2D}}=\sqrt{\hbar/m\omega_{0}}=\sigma$. With decreasing $\sigma$, the 2D features in the scattering of two atoms become pronounced~\cite{petrov2000,petrov2001,zhou2010}, and in the limit $\sigma\ll a_{\text{3D}}$, Eq.~(\ref{Gh}) becomes independent of the value of $a_{\text{3D}}$, and a regime of purely 2D scattering is achieved, with Eq.~(\ref{Gh}) reducing to the coupling constant of a purely 2D Bose gas $g_{\text h}\rightarrow g_{\text{2D}}$
\begin{eqnarray}
g_{\text{2D}} & = & \frac{4\pi\hbar^{2}}{m}\frac{1}{\ln\left(1/n_{\text{2D}}a_{\text{2D}}^{2}\right)}.\label{G2D}
\end{eqnarray}
In above, the logarithmic dependence on the gas parameter $n_{\text{2D}}a_{\text{2D}}$ is unique of the 2D geometry \cite{zhou2010}. 

(iii) Moreover,  the s-wave finite-range interaction coupling constant in Eq.~(\ref{EFunctional}) can be deduced into the 2D form of $\frac{g_2\tilde{q}^2}{2}=\frac{2g_2m}{\hbar^2}\times \left[\frac {\hbar^2k_\perp^2}{2m(\sqrt{8\pi}\sigma/d)}\right]$. We remark that the emphasis and value of the present work is to study the effect of dimensional crossover induced by an optical lattice on the finite-range interaction. The following analysis focuses on the anisotropic 3D and 2D geometry behavior of the ground-state energy given in Eq.~(\ref{GSEC}) and the quantum depletion described in Eq.~(\ref{Qdeple}). 

Now, we are ready to explore the nonuniversal behaviors of EOS of the model system along the dimensional crossover from 3D to quasi-2D based on Eqs.~(\ref{GSE}) and~(\ref{Qdeple}).  Eq.~(\ref{Fx}) has been integrated numerically, and the result are shown in Fig.~\ref{1}(a).

In the limit $x=2t/g_en_0\gg1$, corresponding to the anisotropic 3D regime, we find that the function of $f(x)$ in Eq.~(\ref{Fx}) approaches the asymptotic law of  $f(x)\simeq32/15\lambda^{2}\sqrt{x}$,which has been plotted into Fig.~\ref{1}(a) with the solid-point curves at the fitting range $x\in [1.72,4.48]$. Introducing the effective mass $m^{*}=\hbar^{2}/2td^{2}$  associated with the band, Eq.~(\ref{GSEC}) takes the asymptotic form~\cite{orso2006,zhou2010}
\begin{eqnarray}
\frac{E_{\text g}}{V}\!=\!\frac{\tilde{g}_{3D}n^{2}}{2}\left[1\!+\!\frac{128}{15(1\!+\!\frac{4m\mu}{\hbar^{2}}\frac{g_{2}}{g_{0}})^{2}}\sqrt{\frac{m^{*}}{m}}\sqrt{\frac{n\tilde{a}_{\text{3D}}^{3}}{\pi}}\right], 
\label{GSE3D}
\end{eqnarray}
where the second term of Eq.~(\ref{GSE3D}) corresponds to the generalized LHY correction in the presence of the optical lattice and finite-range interaction. It should be noted that Eq.~(\ref{GSE3D}) can be simplified to yield the respective outcomes presented in Ref.~\cite{zhou2010} when the finite-range interaction represented by $g_2$ vanishes; moreover, Eq.~(\ref{GSE3D}) aligns with the outcomes presented in Ref.~\cite{cappellaro2017} in the absence of an optical lattice, provided that $\tilde{a}_{3D}$ is substituted with $a_{s}$. We define $\lambda=1+\frac{4m\mu}{\hbar^{2}}\frac{g_{2}}{g_{0}}$, which characterizes the effects of the finite-range interaction. Then we select different values of $\lambda$ and plot the corresponding outcomes from Eq.~({\ref{GSE3D}}) as the solid-point curves in Fig.~\ref{1}(a). It is evident that the ground state energy we calculate from Eq.~(\ref{GSE}) can be extrapolated to yield the anisotropic 3D results presented in Eq.~(\ref{GSE3D}).

In the opposite 2D regime corresponding to $x\ll1$, the $f(x)$ in Eq.~(\ref{Fx}) approaches the asymptotic law of $f(x)\simeq\pi/4\lambda^{3/2}-\pi\ln[2\lambda(x+2(1.07/\lambda)^{4}t_{2})]/2\lambda^{3/2}$,  as shown by the empty-point curved in Fig.~\ref{1}(a) at the fitting range $x\in [0.01, 0.58]$. In this limit, Eq.~(\ref{GSEC} yields the ground-state energy of a 2D Bose gas with the consideration of the finite-range interaction
\begin{eqnarray}
\frac{E_{\text{g2D}}}{L^{2}} & \simeq & \frac{1}{2}g_{\text{2D}}n_{\text{2D}}^{2}\Big\{1+\frac{mg_{\text{2D}}}{8\pi\hbar^{2}\lambda^{3/2}}\nonumber \\
&  & -\frac{mg_{\text{2D}}}{4\pi\hbar^{2}\lambda^{3/2}}\ln\left[\lambda\left(\frac{4(t+B2g_{2}n_{0}t_1)}{g_{\text{2D}}n_{\text{2D}}}\right)\right]\Big\}, 
\label{GSE2D}
\end{eqnarray}
where $L^{2}$ is the surface area of the gas, $n_{\text{2D}}=n_{0}d$ is the surface density, and $B=\left(\frac{1.07}{\lambda}\right)^{4}g_{\text{2D}}n_{\text{2D}}$. In the absence of an optical lattice, Eq.~(\ref{GSE2D}) reproduces the respective ground state energy results for a 2D dilute Bose gas with finite-range interaction presented in Ref.~\cite{tononi2018}. It should be noted that the nonuniversal parameter $\lambda$ also appears within the logarithmic term. Subsequently, we select different values of $\lambda$ and plot the corresponding outcomes from Eq.~(\ref{GSE2D}) as the empty-point curves in Fig.~\ref{1}(a). These curves demonstrate that the ground state energy we calculate from Eq.~(\ref{GSE}) can be extrapolated to yield the anisotropic 2D results presented in Eq.~(\ref{GSE2D}).

In a similar fashion, we analyze the asymptotic behavior of quantum depletion using Eq.~(\ref{Qdeple}). In the limit where $x\gg1$ corresponds to the anisotropic 3D regime, we discover that $h(x)\simeq2(\lambda-2)/3\lambda^{2}x^{1/2}$, as depicted by the solid-point curves in Fig.~\ref{1}(b) at the fitting range $x\in [1.72,4.48]$. Subsequently, by inserting the asymptotic law of $h(x)$ into Eq.~(\ref{Qdeple}), one can directly derive the analytical expression for quantum depletion in the 3D limit
\begin{eqnarray}
\frac{\Delta N}{N}|_{\text{3D}} & \simeq & \frac{8}{3\pi^{1/2}}\sqrt{\frac{m^{*}}{m}}(n\tilde{a}_{\text{3D}}^{3})^{1/2}\nonumber \\
 &  & -64C^{2}\pi^{1/2}\sqrt{\frac{m^{*}}{m}}\frac{r_{s}}{\tilde{a}_{\text{3D}}}(n\tilde{a}_{\text{3D}}^{3})^{3/2},\label{QD3D}
\end{eqnarray}
with $C=d\int_{-d/2}^{d/2}\omega^{4}(u)du\simeq d/\sqrt{2\pi}\sigma$.

Note that the quantum depletion in Eq.~(\ref{QD3D}) generalizes the result presented in Ref.~\cite{tononi2018} to include the introduction of an optical lattice. Furthermore, in the opposite limit where $x\ll1$ , corresponding to the 2D limit, the function of $h(x)$  approaches
\begin{eqnarray}
h(x)& \simeq& \frac{\pi}{2\lambda^{3/2}}\left[\frac{\lambda+1}{2}+\frac{\lambda-1}{2}\left(\ln\left[\left(\lambda x\right)^{\alpha}\right]-\frac{C_{1}}{C_{2}+\ln\left[\lambda\right]}\right)\right],
\nonumber\\
\label{H2D}
\end{eqnarray}
with $\alpha=0.126/(\lambda-0.999)$, $C_1=1.715$ and $ C_2=\ln 1.014$, and Eq.~(\ref{H2D}) has been plotted into Fig.~\ref{1}(b) with the empty-point curves at the fitting range $x\in [0.02, 0.58]$. When there is no finite-range interaction, by setting $\lambda=1$, Eq.~(\ref{H2D}) recovers the corresponding result presented in Ref.~\cite{zhou2010}. The quantum depletion in the 2D case is then given by
\begin{eqnarray}
&&\frac{\Delta N}{N}|_{\text{2D}}  \simeq  \frac{1}{\lambda^{3/2}\ln\left(1/n_{\text{2D}}a_{\text{2D}}^{2}\right)}\Big\{\frac{\lambda+1}{2}\nonumber \\
 &  &+\frac{\lambda-1}{2}\left(\ln\left[\left(\frac{2\lambda t}{g_{\text{2D}}n_{\text{2D}}}\right)^{\alpha}\right]-\frac{C_{1}}{C_{2}+\ln\left[\lambda\right]}\right)\Big\}.
\label{Deple2D}
\end{eqnarray}
 Our result in Eq.~(\ref{Deple2D}) shows good agreement with the one presented in Ref.~\cite{tononi2018} regarding the quantum depletion of a purely 2D Bose gas with finite-range interaction.

By choosing the different values of finite-range interaction of $\lambda$, we have plotted the results from both Eqs.~(\ref{QD3D}) and~(\ref{Deple2D}) in Fig.~\ref{1}(b).  One can find that our calculated quantum depletion from Eq.~(\ref{Qdeple}) accurately extrapolate to the 3D results of Eq.~(\ref{QD3D}) in the limit of $x\gg 1$ (as seen in the solid-point curves in Fig.~\ref{1}(b)) and to the 2D results of Eq.~(\ref{Deple2D}) in the limit where $x\ll 1$ (see the empty-point curves in Fig.~\ref{1}(b)). 

\section{CONCLUSION and outlook}\label{CCS}

The emphasis and value of this work lie in visualizing the finite-range effects on the system's ground-state properties under the dimensional crossover from 3D to quasi-2D. Moreover, the Bogoliubov approximation used in our calculation should be justified a posteriori by estimating the quantum depletion. For an optically-trapped Bose gas along the dimensional crossovers, we can estimate the quantum depletion $(N-N_0)/N$ with the help of Fig.~\ref{1}(b). In typical experiments with an optically trapped Bose-Einstein condensate (BEC), the relevant parameters are $n=3\times10^{13}~\text{cm}^{-3}$, $d=430~\text{nm}$, $a_{\text{3D}}=5.4~\text{nm}$, and $d/\sigma\sim1$~\cite{bloch2008}. Thus, the quantum depletion in Eq.~(\ref{Qdeple}) is evaluated as $(N-N_0)/N\sim 0.0031\times h(x)$, with $h(x)$ shown in Fig.~\ref{1}(b). Therefore, the Bogoliubov approximation is valid~\cite{Xu2006}. The experimental realization of our scenario involves controlling three parameters: the strength of the optical lattice $s$, the s-wave scattering length $a_{\text{s}}$, and the effective range $r_{\text{e}}$. All these parameters are highly controllable using state-of-the-art technologies: The depth of an optical lattice $s$ can be changed from 0$E_{\text R}$ to 32$E_{\text R}$~\cite{Greiner2002}, and both $a_{\text{s}}$ and $r_{\text{e}}$ can be controlled by the dark-state method~\cite{Liu2014,WuPRA2012}. We hope the predicted results can be observed in future experiments.

 We mention that there exists a scaling symmetry under the transformation $\bf{r}\rightarrow \lambda \bf{r}$ for a pure 2D Bose gas interacting only via $g_0\delta(\bf{r})$ potential.  Associated with this scale invariance is an underlying symmetry SO(2,1) symmetry~\cite{Pitaevskii1997} and a universal frequency belonging with breathing modes. As such,  the broken SO(2,1) symmetry by the finite-range and dimensional effects, inducing a frequency shift in breathing mode, provide sensitive measurements of quantum many-body effects. Along this research line, the finite-range correction to the universal breathing mode in quasi-2D Fermi gases has been investigated in connection with quantum anomalies~\cite{Hu2019,Yin2020}. In the context of a Bose gas, Ref. \cite{hu2011} has studied how the quasi-two-dimensionality affects the breathing mode and the scale invariance. In this sense,  observing the derivation of  this universality of the breathing mode in Ref. \cite{Jalm2019} presents an important step in revealing the interplay between dimensionality and quantum fluctuations in quasi-2D. To our best knowledge, there is no related work studying how the finite-range correction affects the breathing mode and the scale invariance in bosonic systems, which needs to solve the hydrodynamic equations. In order to maintain the self-consistency of the current work, we will leave this interesting question for the future research. 

In summary, the purpose of this work is to investigate nonuniversal corrections to EOS for an optically trapped Bose gas along the dimensional crossover from 3D to 2D, incorporating the finite-range effects of the interatomic potential. Capitalizing on the characteristic dimensional crossover properties, the results obtained in the quasi-2D regime enable us to derive analytical expressions for the ground-state energy and quantum depletion of an effectively pure 2D Bose gas with finite-range interaction. Our analysis can also demonstrates that ground-state properties are logarithmically dependent on nonuniversal parameters in systems with reduced dimensionality.

We thank Kangkang Li and Ying Hu for stimulating discussions. This work was supported by the National Natural Science Foundation of China (Nos. 12074344), the Zhejiang Provincial Natural Science Foundation (Grant Nos. LZ21A040001) and the key projects of the Natural Science Foundation of China (Grant No. 11835011).
\newpage
\onecolumngrid
\section*{Appendixes}
\appendix
\section{ Detailed Derivation of Action functional of Eq.~(\ref{EFunctional})}\label{AppendixA}

In Appendix \ref{AppendixA},  we plan to give the detailed derivations of Eq.~(\ref{EFunctional}) starting from Eq.~(\ref{Lg}). Here we will mainly derivate
the interacting terms, while the non-interacting terms $\sum_{{\bf k},n}\psi_{{\bf k},n}^{*}(-i\hbar\omega_{n}+\varepsilon_{{\bf k}}^{0}-\mu)\psi_{{\bf k},n}^{*}$
have been discussed in \cite{zhou2010,hu2013}. As such, we will discuss the contact interacting term and finite-range interacting term respectively.

After expanding the complex bosonic field to Eq.~(\ref{CPF}), the contact interacting term $S_{\text{ci}}$ of the action functional can be written as
\begin{eqnarray}
	\frac{S_{\text{ci}}}{\hbar\beta V}  &=& \frac{1}{\hbar\beta V}\int_{0}^{\hbar\beta}d\tau\int d^{3}\mathbf{r}\frac{g_{0}}{2}\left|\psi\left(\mathbf{r},\tau\right)\right|^{4}\nonumber\\
	& = & \frac{1}{\hbar\beta V}\sum_{\substack{\mathbf{k_{1},k_{2},k_{3},k_{4}},l\nonumber\\
			n_{1},n_{2},n_{3},n_{4}
		}
	}\frac{g_{0}}{2}\psi_{\mathbf{k_{1}},n_{1}}^{*}\psi_{{\bf k_{2}},n_{2}}\psi_{\mathbf{k_{3}},n_{3}}^{*}\psi_{{\bf k_{4}},n_{4}} \int_{0}^{\hbar\beta}e^{-i(\omega_{n1}-\omega_{n2}+\omega_{n3}-\omega_{n4})\tau}d\tau\nonumber \\
	&& \int e^{i(k_{y1}-k_{y2}+k_{y3}-k_{y4})y}dy \int e^{i(k_{z1}-k_{z2}+k_{z3}-k_{z4})z}dz \int\omega^{4}(x-ld)dxe^{i(k_{x1}-k_{x2}+k_{x3}-k_{x4})ld}\nonumber\\
	&=&  \sum_{\substack{\mathbf{k_{1},k_{2},q}\\
			n_{1},n_{2},m
		}
	}\frac{g_{\text e}}{2}\psi_{\mathbf{k_{2}+q},n_{2}+m}^{*}\psi_{\mathbf{k_{1}-q},n_{1}-m}^{*}\psi_{{\bf k_{1}},n_{1}}\psi_{{\bf k_{2}},n_{2}},\label{AC1}
\end{eqnarray}
with $g_{\text e}=g_{0}d/\sqrt{2\pi}\sigma $.

We proceed to derive the finite-range interacting term of Eq.~(\ref{EFunctional}). In more details, we will derivate the action functional in lattice direction ($x$ direction) and $y-z$ directions respectively.
First, we focus on the action functional in $y$ and $z$ directions. Plugging the expanded Bosonic field Eq.~(\ref{CPF}) into Eq.~(\ref{Lg}), we can obtain 
\begin{eqnarray}
	\frac{S_{\text{fiyz}}}{\hbar\beta V}  &=&  -\frac{1}{\hbar\beta V}\int_{0}^{\hbar\beta}d\tau\int d^{3}\mathbf{r}\frac{g_{2}}{2}\left|\psi\left(\mathbf{r},\tau\right)\right|^{2}(\partial_{y}^{2}+\partial_{z}^{2})\left|\psi\left(\mathbf{r},\tau\right)\right|^{2}\nonumber\\
	& =&  -\frac{1}{\hbar\beta V}\int_{0}^{\hbar\beta}d\tau\int dxdydz\sum_{\substack{\mathbf{k_{1},k_{2},k_{3},k_{4}},l\nonumber\\
			n_{1},n_{2},n_{3},n_{4}
		}
	}e^{i(k_{x1}-k_{x2}+k_{x3}-k_{x4})ld}e^{-i(\omega_{n1}-\omega_{n2}+\omega_{n3}-\omega_{n4})\tau}\omega^{4}(x-ld)\nonumber \\
	&& \frac{g_{2}}{2} \psi_{\mathbf{k_{1}},n_{1}}^{*}\psi_{{\bf k_{2}},n_{2}}e^{i\left(\left(k_{y1}-k_{y2}\right)y+\left(k_{z1}-k_{z2}\right)z\right)}(\partial_{y}^{2}+\partial_{z}^{2})\psi_{\mathbf{k_{3}},n_{3}}^{*}\psi_{{\bf k_{4}},n_{4}}e^{i\left(\left(k_{y3}-k_{y4}\right)y+\left(k_{z3}-k_{z4}\right)z\right)}\nonumber \\
	&=&  \sum_{\substack{\mathbf{k_{1},k_{2},q}\\
			n_{1},n_{2},m
		}
	}\frac{g_{2}(q_{z}^{2}+q_{y}^{2})d}{2\sqrt{2\pi}\sigma}\psi_{\mathbf{k_{2}+q},n_{2}+m}^{*}\psi_{\mathbf{k_{1}-q},n_{1}-m}^{*}\psi_{{\bf k_{1}},n_{1}}\psi_{{\bf k_{2}},n_{2}}. \label{AC2}
\end{eqnarray}
Second, we can derive the action functional in the lattice direction. 
\begin{eqnarray}
	\frac{S_{\text{fix}}}{\hbar\beta V}  &=&  -\frac{1}{\hbar\beta V}\int_{0}^{\hbar\beta}d\tau\int d^{3}\mathbf{r}\frac{g_{2}}{2}\left|\psi\left(\mathbf{r},\tau\right)\right|^{2}\partial_{x}^{2}\left|\psi\left(\mathbf{r},\tau\right)\right|^{2}\nonumber\\
    & = &
  -\int_{0}^{\hbar\beta}d\tau\int dxdydz\sum_{\substack{\mathbf{k_{1},k_{2},k_{3},k_{4}}\\
  		n_{1},n_{2},n_{3},n_{4}}}\sum_{l_{1},l_{2},l_{3},l_{4}}\frac{g_{2}}{2}\psi_{\mathbf{k_{1}},n_{1}}^{*}\psi_{{\bf k_{2}},n_{2}}e^{i(k_{y1}-k_{y2})y}e^{i(k_{z1}-k_{z2})z}\omega(x-l_{1}d)\omega(x-l_{2}d)\nonumber \\
  &&e^{i(k_{1x}l_{1}-k_{2x}l_{2})}e^{-i(\omega_{n1}-\omega_{n2}+\omega_{n3}-\omega_{n4})\tau}\partial_{x}^{2}\psi_{\mathbf{k_{3}},n_{1}}^{*}\psi_{{\bf k_{4}},n_{2}}e^{i(k_{y3}-k_{y4})y}e^{i(k_{z3}-k_{z4})z}\omega(x-l_{3}d)\omega(x-l_{4}d)e^{i(k_{3x}l_{3}-k_{4x}l_{4})}\nonumber\\
	& =&  -\frac{g_{2}}{2}\frac{1}{L_{x}}\sum_{\substack{\mathbf{k_{1},k_{2},q}\\
			n_{1},n_{2},m
		}
	}\sum_{l_{1},l_{2}}	 e^{iq_{x}d\left(l_{1}-l_{2}\right)}\psi_{\mathbf{k_{2}+q},n_{2}+m}^{*}\psi_{\mathbf{k_{1}-q},n_{1}-m}^{*}\psi_{{\bf k_{1}},n_{1}}\psi_{{\bf k_{2}},n_{2}}\int dx\omega^{2}(x-l_{1}d)\partial_{x}^{2}\omega^{2}(x-l_{2}d) \nonumber\\ 
   & = & -\frac{1}{d}\frac{g_{2}}{2}\sum_{\substack{\mathbf{k_{1},k_{2},q}\nonumber\\
			n_{1},n_{2},m
		}
	}\psi_{\mathbf{k_{2}+q},n_{2}+m}^{*}\psi_{\mathbf{k_{1}-q},n_{1}-m}^{*}\psi_{{\bf k_{1}},n_{2}}\psi_{{\bf k_{2}},n_{2}}\left(\int dx\omega^{2}(x)\partial_{x}^{2}\omega^{2}(x)+2\cos q_{x}d\int dx\omega^{2}(x)\partial_{x}^{2}\omega^{2}(x-d)\right)\nonumber \\
	& =  &\frac{g_{2}}{2}\sum_{\substack{\mathbf{k_{1},k_{2},q}\\
	n_{1},n_{2},m}}2t_{1}(1-\cos q_{x}d)\psi_{\mathbf{k_{2}+q},n_{2}+m}^{*}\psi_{\mathbf{k_{1}-q},n_{1}-m}^{*}\psi_{{\bf k_{1}},n_{1}}\psi_{{\bf k_{2}},n_{2}},\label{AC3}
\end{eqnarray}
with $t_1=-\frac{1}{2d}\int dx\omega^{2}(x)\partial_{x}^{2}\omega^{2}(x)=\frac{1}{d}\int dx\omega^{2}(x)\partial_{x}^{2}\omega^{2}(x-d)$. Note that Eqs. (\ref{AC1}), (\ref{AC2}) and (\ref{AC3}) correspond to the terms in the second line of Eq. (\ref{EFunctional}).

\section{Remove power ultraviolet divergences of Eq. (\ref{GSE})}\label{AppendixB}
In Appendix \ref{AppendixB}, we plan to give the detailed derivations of the crucial regularizing terms on the second line of Eq. (\ref{GSE}).  The original form the ground state energy reads as
$\frac{E_{\text g}}{V} =\frac{1}{2}g_{\text e}n_{0}^{2}+\frac{1}{2V}\sum_{\mathbf{k}\ne0} E_{\mathbf{k}} $, which is divergent in the large $\bf{k}$ limit.  Following the procedure of avoiding ultraviolet divergences in Ref. \cite{braaten1997}, the ground state energy can be written as follows
\begin{eqnarray}
\frac{E_{\text g}}{V} =\frac{1}{2}g_{\text e}n_{0}^{2}+\frac{1}{2V}\sum_{\mathbf{k}\ne0}\left[ E_{\mathbf{k}}-\lim_{\bf {k}\rightarrow \infty}E_{\mathbf{k}}\right],\label{AppenG}
\end{eqnarray}
with
\begin{eqnarray}
	\lim_{{\bf {k}\rightarrow\infty}}E_{{\bf k}}&=&\lim_{{\bf {k}\rightarrow\infty}}\sqrt{\left(\frac{\hbar^{2}k_{\perp}^{2}}{2m}+2t\left(1-\cos k_{x}d\right)\right)\left(\frac{\hbar^{2}k_{\perp}^{2}}{2m/\left(1+\frac{4m\mu}{\hbar^{2}}\frac{g_{2}}{g_{0}}\right)}+\frac{\hbar^{2}\left(1-\cos k_{x}d\right)}{\hbar^{2}/2\left(t+2g_{2}n_{0}t_{1}\right)}+2\mu\right)}\nonumber\!\!\!\!\\
	&=&\sqrt{\left(\frac{\hbar^{2}k_{\perp}^{2}}{2m}+2t\left(1-\cos k_{x}d\right)\right)\left(\frac{\hbar^{2}k_{\perp}^{2}}{2m/\left(1+\frac{4m\mu}{\hbar^{2}}\frac{g_{2}}{g_{0}}\right)}+\frac{\hbar^{2}\left(1-\cos k_{x}d\right)}{\hbar^{2}/\left[2\left(t+2g_{2}n_{0}t_{1}\right)\right]}\right)}\nonumber\\
	&&+\frac{g_{e}n_{0}\sqrt{\frac{\hbar^{2}k_{\perp}^{2}}{2m}+2t\left(1-\cos k_{x}d\right)}}{\sqrt{\frac{\hbar^{2}k_{\perp}^{2}}{2m/\left(1+\frac{4m\mu}{\hbar^{2}}\frac{g_{2}}{g_{0}}\right)}+\frac{\hbar^{2}\left(1-\cos k_{x}d\right)}{\hbar^{2}/\left[2\left(t+2g_{2}n_{0}t_{1}\right)\right]}}}-\frac{\left(g_{e}n_{0}\right)^{2}\sqrt{\frac{\hbar^{2}k_{\perp}^{2}}{2m}+2t\left(1-\cos k_{x}d\right)}}{2\left(\frac{\hbar^{2}k_{\perp}^{2}}{2m/\left(1+\frac{4m\mu}{\hbar^{2}}\frac{g_{2}}{g_{0}}\right)}+\frac{\hbar^{2}\left(1-\cos k_{x}d\right)}{\hbar^{2}/[2\left(t+2g_{2}n_{0}t_{1}\right)]}\right)^{3/2}}+O\left(\frac{1}{k_{\perp}^{4}}\right)\label{CT}
\end{eqnarray}
Plugging Eq. (\ref{CT}) into Eq. (\ref{AppenG}), we can obtain Eq. (\ref{GSE}) in the main text.

After removing the power divergences in Eq. (\ref{GSE}), we can replace the summation with an integral in continuum limit and deduce Eq. (\ref{GSEC}) as follows
\begin{eqnarray}
&&\frac{E_{\text{g}}}{V}  =  \frac{1}{2}g_{\text{e}}n_{0}^{2}+\frac{1}{2V}\sum_{{\bf {k}\ne0}}E_{{\bf k}}-E_{{\bf k}}^{0}\nonumber\\
& =&  \frac{1}{2}g_{\text{e}}n_{0}^{2}+\frac{1}{2V}\frac{1}{\Delta k_{x}\Delta k_{y}\Delta k_{z}}\int dk_{x}\int dk_{\perp}2\pi k_{\perp}\Big\{ E_{\mathbf{k}}-\sqrt{\Big[\frac{\hbar^{2}k_{\perp}^{2}}{2m}+2t(1-\cos k_{x}d)\Big]\Big[\frac{\hbar^{2}k_{\perp}^{2}}{2m/(1+\frac{4m\mu}{\hbar^{2}}\frac{g_{2}}{g_{0}})}+\frac{\hbar^{2}(1-\cos k_{x}d)}{\hbar^{2}/[2\left(t+2g_{2}n_{0}t_{1}\right)]}\Big]}\nonumber\\
&  & -{\frac{g_{e}n_{0}\sqrt{\frac{\hbar^{2}k_{\perp}^{2}}{2m}+2t\gamma}}{\sqrt{\frac{\hbar^{2}k_{\perp}^{2}}{2m/(1+\frac{4m\mu}{\hbar^{2}}\frac{g_{2}}{g_{0}})}+\frac{\hbar^{2}(1-\cos k_{x}d)}{\hbar^{2}/\left[2\left(t+2g_{2}n_{0}t_{1}\right)\right]}}}+\frac{(g_{e}n_{0})^{2}\sqrt{\frac{\hbar^{2}k_{\perp}^{2}}{2m}+2t(1-\cos k_{x}d)}}{2\left(\frac{\hbar^{2}k_{\perp}^{2}}{2m/(1+\frac{4m\mu}{\hbar^{2}}\frac{g_{2}}{g_{0}})}+\frac{\hbar^{2}(1-\cos k_{x}d)}{\hbar^{2}/[2\left(t+2g_{2}n_{0}t_{1}\right)]}\right)^{3/2}}\Big\}}\nonumber\\
& = & \frac{1}{2}g_{\text{e}}n_{0}^{2}+\frac{1}{2(2\pi)^{2}d}\int dk_{x}^{\prime}\int dk_{\perp}k_{\perp}\Big\{ E_{\mathbf{k}}-\sqrt{\Big[\frac{\hbar^{2}k_{\perp}^{2}}{2m}+2t(1-\cos k_{x}^{\prime})\Big]\Big[\frac{\hbar^{2}k_{\perp}^{2}}{2m/(1+\frac{4m\mu}{\hbar^{2}}\frac{g_{2}}{g_{0}})}+\frac{\hbar^{2}(1-\cos k_{x}^{\prime})}{\hbar^{2}/[2\left(t+2g_{2}n_{0}t_{1}\right)]}\Big]}\nonumber\\
&  & {-\frac{g_{e}n_{0}\sqrt{\frac{\hbar^{2}k_{\perp}^{2}}{2m}+2t(1-\cos k_{x}^{\prime})}}{\sqrt{\frac{\hbar^{2}k_{\perp}^{2}}{2m/(1+\frac{4m\mu}{\hbar^{2}}\frac{g_{2}}{g_{0}})}+\frac{\hbar^{2}(1-\cos k_{x}^{\prime})}{\hbar^{2}/\left[2\left(t+2g_{2}n_{0}t_{1}\right)\right]}}}+\frac{(g_{e}n_{0})^{2}\sqrt{\frac{\hbar^{2}k_{\perp}^{2}}{2m}+2t(1-\cos k_{x}^{\prime})}}{2\left(\frac{\hbar^{2}k_{\perp}^{2}}{2m/(1+\frac{4m\mu}{\hbar^{2}}\frac{g_{2}}{g_{0}})}+\frac{\hbar^{2}(1-\cos k_{x}^{\prime})}{\hbar^{2}/[2\left(t+2g_{2}n_{0}t_{1}\right)]}\right)^{3/2}}\Big\}}\nonumber\\
& = & \frac{1}{2}g_{\text{e}}n_{0}^{2}+\frac{1}{4(2\pi)^{2}d}\frac{2m}{\hbar^{2}}\int dk_{x}^{\prime}\int d\frac{\hbar^{2}k_{\perp}^{2}}{2m}\Big\{ E_{\mathbf{k}}-\sqrt{\Big[\frac{\hbar^{2}k_{\perp}^{2}}{2m}+2t(1-\cos k_{x}^{\prime})\Big]\Big[\frac{\hbar^{2}k_{\perp}^{2}}{2m/(1+\frac{4m\mu}{\hbar^{2}}\frac{g_{2}}{g_{0}})}+\frac{\hbar^{2}(1-\cos k_{x}^{\prime})}{\hbar^{2}/[2\left(t+2g_{2}n_{0}t_{1}\right)]}\Big]}\nonumber \\
&  & {-\frac{g_{e}n_{0}\sqrt{\frac{\hbar^{2}k_{\perp}^{2}}{2m}+2t(1-\cos k_{x}^{\prime})}}{\sqrt{\frac{\hbar^{2}k_{\perp}^{2}}{2m/(1+\frac{4m\mu}{\hbar^{2}}\frac{g_{2}}{g_{0}})}+\frac{\hbar^{2}(1-\cos k_{x}^{\prime})}{\hbar^{2}/\left[2\left(t+2g_{2}n_{0}t_{1}\right)\right]}}}+\frac{(g_{e}n_{0})^{2}\sqrt{\frac{\hbar^{2}k_{\perp}^{2}}{2m}+2t(1-\cos k_{x}^{\prime})}}{2\left(\frac{\hbar^{2}k_{\perp}^{2}}{2m/(1+\frac{4m\mu}{\hbar^{2}}\frac{g_{2}}{g_{0}})}+\frac{\hbar^{2}(1-\cos k_{x}^{\prime})}{\hbar^{2}/[2\left(t+2g_{2}n_{0}t_{1}\right)]}\right)^{3/2}}\Big\}}\nonumber
\end{eqnarray}
\begin{eqnarray}
& = & \frac{1}{2}g_{\text{e}}n_{0}^{2}+\frac{1}{2(2\pi)^{2}d}\frac{m}{\hbar^{2}}\int dk_{x}^{\prime}\int dK\Big\{ E_{\mathbf{k}}-\sqrt{\Big[K+2t\gamma\Big]\Big[\lambda K+2\left(t+2g_{2}n_{0}t_{1}\right)\gamma\Big]}\nonumber\\
&  & {-\frac{g_{e}n_{0}\sqrt{K+2t\gamma}}{\sqrt{\lambda K+2\left(t+2g_{2}n_{0}t_{1}\right)\gamma}}+\frac{(g_{e}n_{0})^{2}\sqrt{K+2t\gamma}}{2\left(\lambda K+2\left(t+2g_{2}n_{0}t_{1}\right)\gamma\right)^{3/2}}\Big\}}\nonumber\\
& = & \frac{1}{2}g_{\text{e}}n_{0}^{2}+\frac{(g_{e}n_{0})^{2}}{2(2\pi)^{2}d}\frac{m}{\hbar^{2}}\int dk_{x}^{\prime}\int dk\Big\{ E_{\mathbf{k}}-\sqrt{\Big[k+x\gamma\Big]\Big[\lambda k+(x+2t_{2})\gamma\Big]} {-\frac{\sqrt{k+x\gamma}}{\sqrt{\lambda k+(x+2t_{2})\gamma}}+\frac{\sqrt{k+x\gamma}}{2\left(\lambda k+(x+2t_{2})\gamma\right)^{3/2}}\Big\}}\nonumber\\
& = & \frac{1}{2}g_{\text{e}}n_{0}^{2}+\frac{(g_{e}n_{0})^{2}}{(2\pi)^{2}d}\frac{m}{\hbar^{2}}f(x),
\end{eqnarray}
{ with $K=\hbar^2 k_{\perp}^2/2m$, $k=K/g_{e}n_0$}.
\twocolumngrid
\bibliography{xyref}
\end{document}